\documentclass[twocolumn,aps,showpacs,prb,tightenlines,amsmath,amssymb]{revtex4}
\usepackage{graphicx} 
\usepackage{amssymb}
\usepackage{dcolumn}
\usepackage{amsmath}
\usepackage{bm}

\usepackage{colordvi}
\newcommand{\bgreek}[1]{\mbox{\boldmath$#1$\unboldmath}}

\begin{document}

\title{Hole spin relaxation and coefficients in Landau-Lifshitz-Gilbert
  equation in ferromagnetic GaMnAs} 

\author{K. Shen}
\affiliation{Hefei National Laboratory for Physical Sciences at
Microscale and Department of Physics,
University of Science and Technology of China, Hefei,
Anhui, 230026, China}
\author{M. W. Wu}
\thanks{Author to  whom correspondence should be addressed}
\email{mwwu@ustc.edu.cn.}
\affiliation{Hefei National Laboratory for Physical Sciences at
Microscale and Department of Physics,
University of Science and Technology of China, Hefei,
Anhui, 230026, China}

\date{\today}

\begin{abstract}
We investigate the temperature dependence of the
coefficients in the Landau-Lifshitz-Gilbert equation in ferromagnetic GaMnAs by
employing the Zener model. We first calculate the hole spin relaxation time
based on the microscopic kinetic equation. We find that the hole
spin relaxation time is typically several tens femtoseconds and can present a nonmonotonic
temperature dependence due to the variation of the interband spin mixing,
influenced by the temperature related Zeeman
splitting. With the hole spin 
relaxation time, we are able to calculate the coefficients
in the Landau-Lifshitz-Gilbert equation, such as the Gilbert damping,
nonadiabatic spin torque, spin stiffness and vertical spin
stiffness coefficients. We find that the nonadiabatic spin torque coefficient $\beta$
is around $0.1\sim 0.3$ at low temperature, which is consistent with the
experiment [Adam {\em et al.}, Phys. Rev. B {\bf 80}, 193204
(2009)]. As the temperature increases, $\beta$ monotonically increases and 
can exceed one in the vicinity of the Curie temperature. In the low temperature regime with
$\beta<1$, the Gilbert damping
coefficient $\alpha$ increases with temperature, showing good agreement with
the experiments [Sinova {\em et al.}, Phys. Rev. B {\bf 69}, 085209 (2004);
Khazen {\em et al.}, {\em ibid.} {\bf 78}, 
195210 (2008)]. Furthermore, we predict that $\alpha$ decreases with increasing temperature
once $\beta>1$ near the Curie temperature. We also find that the spin stiffness
decreases with increasing temperature, especially near the Curie temperature due to the
modification of the finite $\beta$. Similar to the
Gilbert damping, the vertical spin stiffness coefficient is also found to be
nonmonotonically dependent on the temperature.
\end{abstract}

\pacs{72.25.Rb, 75.50.Pp, 72.25.Dc, 75.30.Gw}


\maketitle

\section{Introduction}\label{introduction}
The ferromagnetic semiconductor, GaMnAs, has been proposed to be a promising
candidate to realize all-semiconductor spintronic devices,\cite{ohno,jungwirth}
where the existence of the ferromagnetic phase in the heavily doped sample
sustains seamless spin injection and detection in normal non-magnetic
semiconductors.\cite{ciorga,pala} One important issue for such applications
lies in the efficiency of the manipulation of the
macroscopic magnetization, which relies on properties of the magnetization dynamics.
Theoretically, the magnetization dynamics can be described by the extended
Landau-Lifshitz-Gilbert (LLG) equation,\cite{gilbert,landau,berger,szhang,shen2,tatara} 
\begin{eqnarray}
  \nonumber
  \dot{\bf n}
  &=&-\gamma {\bf n}\times {\bf  H}_{\rm
    eff}+{\alpha}{\bf n}\times{\dot{\bf n}}
-(1-{\beta}{\bf n}\times)({\bf v}_s\cdot
  \nabla){\bf n}\\
  &&\hspace{-0.1cm}\mbox{}
  -\tfrac{\gamma}{M_d} {\bf n}\times(A_{\rm ss}-A_{\rm ss}^{\rm v}{\bf
    n}\times)\nabla^2{\bf n},
  \label{LLG}
\end{eqnarray}
with ${\bf n}$ and $M_d$ standing for the direction and magnitude of the
magnetization, respectively. ${\bf H}_{\rm eff}$ is the effective magnetic
field and/or the external field. The second term on the right hand side of the
equation is the Gilbert damping torque with $\alpha$ denoting the damping
coefficient.\cite{gilbert,landau} The third one describes the
spin-transfer torque induced by 
the spin current ${\bf v}_s$.\cite{berger,szhang} As reported, the out-of-plane
contribution of the 
spin-transfer torque, measured by the nonadiabatic torque coefficient $\beta$,
can significantly ease the domain wall motion.\cite{berger,szhang} In
Eq.\,(\ref{LLG}), the spin 
stiffness and vertical spin stiffness coefficients are evaluated by $A_{\rm ss}$ and $A_{\rm
  ss}^{\rm v}$ respectively, which are essentially important for the static
structure of the magnetic domain
wall.\cite{shen2} Therefore, for a thorough understanding of properties of the
magnetization dynamics, the exact values of the above
coefficients are required.

In the past decade, the Gilbert damping and nonadiabatic torque coefficients
have been derived via many microscopic approaches, such as the Blotzmann equation,\cite{piechon}
diagrammatic calculation,\cite{kohno1,kohno2} Fermi-surface breathing
model\cite{kunes,steiauf,tserk3} and kinetic spin Bloch equations.\cite{shen1,shen2}
According to these works, the spin lifetime of the carriers was found to be
critical to both $\alpha$ and $\beta$. However, to the best of our knowledge, the
microscopic calculation of the hole spin lifetime in ferromagnetic GaMnAs is
still absent in the literature, which prevents the determination of the
values of $\alpha$ and
$\beta$ from the analytical formulas. Alternatively, Sinova
{\em et al.}\cite{sinova} identified the Gilbert damping from the susceptibility
diagram of the linear-response theory and calculated $\alpha$ as function of
the quasiparticle lifetime and the hole
density. Similar microscopic calculation on $\beta$ was
later given by Garate {\em et al.}.\cite{garate} In those works, the
quasiparticle lifetime was also treated as a
parameter instead of explicit calculation. Actually, the accurate calculation of
the hole spin and/or quasiparticle lifetime in ferromagnetic GaMnAs is
difficult due to the complex band structure of the valence bands.
In the present work, we employ the microscopic kinetic equation
to calculate the spin lifetime of the hole gas and then evaluate $\alpha$ and $\beta$
in ferromagnetic GaMnAs.
For the velocity of the domain-wall motion due to the spin current, the ratio
$\beta/\alpha$ is an important parameter,
which has attracted much attention.\cite{garate,kohno1,hals} Recently, a huge ratio
($\sim 100$) in nanowire was predicted from the calculation of the scattering matrix
by Hals {\em et  al.}.\cite{hals} By calculating $\alpha$ and $\beta$,
we are able to supply detailed information of this interesting ratio in bulk material.
Moreover, the peak-to-peak ferromagnetic resonance measurement revealed
pronounced temperature and sample preparation dependences of the Gilbert
damping coefficient.\cite{sinova,khazen,qi} For example, in annealed samples,
$\alpha$ can present an increase in
the vicinity of the Curie temperature,\cite{sinova,khazen} which has not
  been studied theoretically in the literature.
Here, we expect to uncover the underlying physics of these features.
In addition, the nonadiabatic torque coefficient $\beta$ in GaMnAs has been
experimentally determined from
the domain-wall motion and quite different values were reported by
different groups, from 0.01 to 0.36,\cite{yaman,adam} which need to be
verified by the microscopic calculation also. Moreover, to the best of our
knowledge, the temperature dependence of $\beta$ has not been studied
theoretically. We will also address this issue in the present work.

In the literature, the spin stiffness in GaMnAs was studied by K\"onig {\em et
  al.},\cite{konig,konig2} who found
that $A_{\rm ss}$ increases with hole density due to the stronger
carrier-mediated interaction between magnetic ions, i.e., $A_{\rm
ss}= N_h/(4m^\ast)$ with $N_h$ and $m^\ast$ being the density and
effective mass of hole gas, separately. However, as shown in our previous work,
the stiffness should be modified as $A_{\rm
ss}\sim N_h/[4m^\ast(1+\beta^2)]$ in ferromagnetic GaMnAs with a finite
$\beta$.\cite{shen2} As a result, $A_{\rm ss}$ as well as the vertical spin
stiffness $A_{\rm ss}^{\rm v}=\beta A_{\rm ss}$ may show a
temperature dependence introduced by $\beta$. This is also a goal of the
present work.

For a microscopic investigation of the hole dynamics, the
valence band structure is required for the description of the occupied carrier states.
In the literature, the Zener model\cite{zener} based on the mean-field theory
has been widely used for itinerant holes in
GaMnAs,\cite{dietl,ablfath,dietl2,cywinski} where the valence bands
split due to the mean-field $p$-$d$ exchange interaction. 
In the present work, we utilize this model to calculate the band structure
with the effective Mn concentration from the
experimental value of the low-temperature saturate magnetization in
GaMnAs. The thermal effect on the band structure is introduced via the temperature
dependence of the magnetization following the Brillouin function. Then we obtain the
hole spin relaxation time by numerically solving the microscopic kinetic 
equations with the relevant hole-impurity and hole-phonon
scatterings. The carrier-carrier scattering is neglected here by considering the
strongly degenerate distribution of the hole gas below the Curie temperature.
We find that the hole spin relaxation time decreases/increases with increasing
temperature in the small/large Zeeman splitting regime, which mainly results from
the variation of the interband spin mixing. Then we study the temperature
dependence of the coefficients in the LLG equation, i.e., $\alpha$, 
$\beta$, $A_{\rm ss}$ and $A_{\rm ss}^{\rm v}$,
by using the analytical formulas derived in our previous works.\cite{shen2,shen1}
Specifically, we find that $\beta$ increases with increasing temperature and
can exceed one in the vicinity of the critical point, resulting in very
interesting behaviors of other coefficients.
For example, $\alpha$ can present an interesting
nonmonotonic temperature dependence with the crossover occurring at $\beta\sim
1$. Specifically, $\alpha$ increases with temperature in the low temperature
regime, which is consistent with the 
experiments. Near the Curie temperature, an opposite temperature dependence
of $\alpha$ is predicted. Similar nonmonotonic behavior is also predicted in the
temperature dependence of $A_{\rm ss}^{\rm v}$. Our results of $\beta$ and
$A_{\rm ss}$ also show good agreement with the experiments.

This work is organized as follows. In Sec.\,II, we setup our model and
lay out the formulism. Then we show the band structure from the Zener model and
the hole spin relaxation time from microscopic kinetic equations in Sec.\,III.
The temperature dependence of the Gilbert damping, nonadiabatic spin torque, spin stiffness and
vertical spin stiffness coefficients are also shown in this section. Finally, we
summarize in Sec.\,IV.

\section{Model and Formulism}\label{model}
In the $sp$-$d$ model, the Hamiltonian of hole gas in GaMnAs is given by\cite{cywinski}
\begin{equation}
  H=H_p+H_{pd},
\label{hami}
\end{equation}
with $H_p$ describing the itinerant
holes. $H_{pd}$ is the $sp$-$d$ exchange coupling. By assuming that the
momentum ${\bf k}$ is still a good quantum number for itinerant hole states,
one employs the Zener model and utilizes
the ${\bf k}\cdot{\bf p}$ perturbation Hamiltonian to describe the
valence band states. Specifically, we take the eight-band Kane
Hamiltonian $H_K({\bf k})$ (Ref.\,\onlinecite{kane}) in the present work. The
$sp$-$d$ exchange 
interaction reads
\begin{equation}
  H_{pd}=-\frac{1}{N_0V}\sum_{l}\sum_{mm^\prime\bf k}J^{mm^\prime}_{\rm ex}{\bf
    S}_{l}\cdot \langle m {\bf k}|\hat{\cal J}|m^\prime{\bf k}\rangle
  c^\dag_{m\bf k}c_{m^\prime\bf k}, 
  \label{eq1}
\end{equation}
with $N_0$ and $V$ standing for the density of cation sites and the volume,
respectively. The cation density $N_0=2.22\times 10^{22}$~cm$^{-3}$. The
eight-band spin operator can be
written as $\hat{\cal J}={(\tfrac{1}{2}\bgreek\sigma)}\oplus {\bf J}_{3/2}\oplus
{\bf J}_{1/2}$, where 
$\frac{1}{2}{\bgreek\sigma}$, ${\bf J}_{3/2}$
and ${\bf J}_{1/2}$ represent the
total angular momentum operators of the conduction band, $\Gamma_8$ valence band
and $\Gamma_7$ valence band, respectively. $J_{\rm ex}^{mm\prime}$ stands for the
matrix element of the exchange coupling, with $\{m\}$ and $\{m^\prime\}$
being the basis defined as the eigenstates of the angular momentum operators
$\hat{\cal J}$. The summation of ``$l$'' is through all localized Mn spins ${\bf
  S}_l$ (at ${\bf r}_l$). 

Then we treat the localized Mn spin in a mean-field approximation and obtain
\begin{equation}
  {\bar H}_{pd}=-x_{\rm eff}\langle {\bf S}\rangle\cdot \left(\sum_{mm^\prime\bf
    k}J^{mm^\prime}_{\rm ex}\langle m {\bf k}|\hat{\cal J}|m^\prime{\bf k}\rangle
  c^\dag_{m\bf k}c_{m^\prime\bf k}\right), 
  \label{eq2}
\end{equation}
where $\langle {\bf S}\rangle$ represents the average spin polarization of
Mn atoms with uncompensated doping density $N_{\rm Mn}=x_{\rm
  eff}{N_0}$. Obviously, ${\bar H}_{pd}$ can be reduced into three blocks as $\hat{\cal
J}$, i.e., ${\bar H}^{mm^\prime}_{pd}({\bf
  k})=\Delta^{mm}{\bf 
  n}\cdot \langle m {\bf k}|\hat{\cal J}|m^\prime{\bf k}\rangle$ with
the Zeeman splitting of the $m$-band $\Delta^{mm}=-x_{\rm eff}S_dJ^{mm}_{\rm
  ex}\frac{M(T)}{M(0)}$. 
Here, ${\bf n}$ is the direction of $\langle {\bf S}\rangle$.
For a manganese ion, the total spin $S_d=5/2$.
The temperature-dependent spontaneous magnetization $M(T)$ can be obtained from the
following equation of the Brillouin function\cite{darby}
\begin{equation}
  B_{S_d}(y)=\frac{S_d+1}{3S_d}\frac{T}{T_c}y,
\label{eq3}
\end{equation}
where $y=\frac{3S_d}{S_d+1}\frac{M(T)}{M(0)}\frac{T_c}{T}$ with $T_c$ being the Curie
temperature. Here,
$B_{S_d}(y)=\frac{2S_d+1}{2S_d}{\rm coth}(\frac{2S_d+1}{2S_d}y)-\frac{1}{2S_d}{\rm
  coth}(\frac{1}{2S_d}y)$.

The Schr\"odinger equation of the single particle Hamiltonian is then
written as 
\begin{equation}
  \big[H_K({\bf k})+{\bar H}_{pd}({\bf k})\big]|\mu,{\bf k}\rangle=E_{\mu{\bf
      k}}|\mu,{\bf k}\rangle. 
\label{eq4}
\end{equation}
One obtains the band structure and wave functions from the diagonalization
scheme. In the presence of a finite Zeeman splitting, the structure of the valence
bands deviates from the parabolic dispersion and becomes strongly anisotropic as
we will show in the next section.
Moreover, the valence bands at Fermi surface are well separated in ferromagnetic
GaMnAs because of the high hole density ($>10^{20}$~cm$^{-3}$) and Zeeman
splitting, suggesting that the Fermi golden rule can be used to 
calculate the lifetime of the quasiparticle states. For example, the contribution of
the hole-impurity scattering on the $\mu$th-band state with energy $\epsilon$ can
be expressed by 
\begin{eqnarray}
  \nonumber
  [\tau_{\mu,p}^{hi}(\epsilon)]^{-1}&=&2\pi\sum_{\nu}\frac{n_i}{D_\mu(\epsilon)}
  \int\frac{d^3k}{(2\pi)^3}\int\frac{d^3q}{(2\pi)^2}
  \delta(\epsilon-\epsilon_{\mu\bf k})\\
  &&\mbox{}\hspace{-1.8cm}\times  
  \delta(\epsilon_{\mu\bf
  k}-\epsilon_{\nu\bf q})
  U_{{\bf k}-{\bf q}}^2
  |\langle \mu {\bf k}|\nu {\bf q}\rangle|^2
  f(\epsilon_{\mu\bf k})[1-f(\epsilon_{\nu\bf q})],
  \label{fermi_golden}
\end{eqnarray}
where $D_\mu(\epsilon)$ stands for the density of states of the $\mu$th
band. $f(\epsilon_{\mu\bf k})$ satisfies the Fermi distribution in the equilibrium
state. The hole-impurity scattering matrix element $U_{\bf 
  q}^2=Z^2e^4/[\kappa_0(q^2+\kappa^2)]^2$ with $Z=1$. $\kappa_0$ and $\kappa$
denote the static dielectric constant and the screening constant under the
random-phase approximation,\cite{schlie} respectively. Similar expression can
also be obtained for the hole-phonon scattering. 

However, it is very complicated to carry out
the multi-fold integrals in Eq.\,(\ref{fermi_golden}) numerically for an
anisotropic dispersion. Also the lifetime of the quasiparticle is not
equivalent to the spin lifetime of the whole system, which is required
to calculate the LLG coefficients according to our previous work.\cite{shen1,shen2}
Therefore, we extend our kinetic spin Bloch equation approach\cite{wu}
to the current system to study the relaxation of the total spin polarization
as follows. By taking into account the finite separation
between different bands, one neglects the interband coherence and
focuses on the carrier dynamics of the 
non-equilibrium population. The microscopic kinetic equation is then given by
\begin{equation}
  \partial_t n_{\mu,{\bf k}}=\partial_t n_{\mu,{\bf k}}\big |^{hi} +\partial_t
  n_{\mu,{\bf k}}\big |^{hp},
  \label{eq5}
\end{equation}
with $n_{\mu,{\bf k}}$ being the carrier occupation factor at the $\mu$th band
with momentum ${\bf k}$. The first and second terms on the right hand side stand
for the hole-impurity and hole-phonon scatterings, respectively. Their
expressions can be written as
\begin{eqnarray}
  \nonumber
  \partial_t n_{\mu,{\bf k}}\big |^{hi}&=&-2\pi n_i \sum_{\nu,{\bf k}^\prime}U^2_{\bf k-
    k'}(n_{\mu\bf k}-n_{\nu\bf k'}){\left|\langle \mu{\bf k}|\nu{\bf k'}
      \rangle\right|^2}\\
  &&\mbox{}\times\delta(E_{\mu\bf k}-E_{\nu\bf k'}),
\label{eq6}
\end{eqnarray}
and 
\begin{eqnarray}
    \nonumber
 \partial_t n_{\mu,{\bf k}}\big |^{hp}&=&-2\pi \sum_{\lambda,\pm,\nu,{\bf
     k}^\prime}|M^\lambda_{\bf k- 
    k'}|^2  \delta(E_{\nu\bf k'}-E_{\mu\bf k}\pm \omega_{\lambda,{\bf
      q}})\\
  &&\hspace{-2.3cm}\mbox{}\times[N^{\pm}_{\lambda, {\bf q}}(1-n_{\nu\bf
    k'})n_{\mu\bf k} -N^{\mp}_{\lambda,{\bf q}}n_{\nu\bf k'}(1-n_{\mu\bf
    k})]{\left|\langle
      \mu{\bf k}|\nu{\bf k'} \rangle\right|^2},
\label{eq7}
\end{eqnarray}
with $N_{\lambda,\bf q}^\pm =[\exp(\omega_{\lambda,\bf
  q}/k_BT)-1]^{-1}+\tfrac{1}{2}\pm\tfrac{1}{2}$. The details of the
hole-phonon scattering elements $|M^\lambda_{\bf q}|^2$ can be found in
Refs.\,\onlinecite{weng,zhou,shen3}.
From an initial condition with a small
non-equilibrium distribution, the temporal evolution of the hole spin
polarization is carried out by 
\begin{equation}
  {\cal J}(t)=\frac{1}{N_h}\sum_{\mu,\bf k}
  \langle\mu{\bf k}|\hat{\cal J}|\mu{\bf k}\rangle n_{\mu,{\bf k}}(t),
\end{equation}
from the numerical solution of Eq.\,(\ref{eq5}). The hole spin
relaxation time can be extracted from the exponential fitting of ${\cal J}$ with
respect to time. One
further calculates the concerned coefficients  
such as $\alpha$, $\beta$, $A_{\rm ss}$ and $A_{\rm ss}^{\rm v}$. 

\section{Numerical Results}\label{results}
In the Zener model, the $sp$-$d$ exchange interaction constants $J_{\rm
  ex}^{mm}$ are important parameters for the band structure.
In the experimental works, the $p$-$d$ exchange coupling constant 
$J^{pp}_{\rm ex}$ was reported to vary from $-1$~eV
to $2.5$~eV, depending on the doping density.\cite{burch,okaba,heim}
In ferromagnetic samples, $J_{\rm ex}^{pp}$ is believed to be negative,
which was demonstrated by theoretical estimation $J^{pp}_{\rm ex}\approx
-0.3$~eV (Ref.\,\onlinecite{tang}).
In our calculation, the antiferromagnetic $p$-$d$ interaction $J_{\rm ex}^{pp}$
is chosen to be $-0.5$~eV or $-1.0$~eV. The ferromagnetic $s$-$d$ exchange
coupling constant
is taken to be $J_{\rm ex}^{ss}=0.2$~eV (Ref.\,\onlinecite{cywinski}). 

Another important quantity for determining the Zeeman splitting is the
macroscopic magnetization or the effective concentration of the Mn atoms.
As deduced from the low-temperature saturate magnetization, only around 50\,\% Mn
atoms can contribute to the ferromagnetic 
magnetization, which has been recognized as the influence of the compensation effect
due to the deep donors (e.g., As antisites) or the formation of sixfold-coordinated centers
defect Mn$^{\rm 6As}$ (Ref.\,\onlinecite{santos}). As only the uncompensated
Mn atoms can supply holes and contribute to the ferromagnetic
magnetization,\cite{esch} one can also estimate the
total hole density from the saturate magnetization.\cite{haghgoo}
However, the density of the itinerant hole can be smaller than the effective Mn
concentration because of the localized effect in such disordered material.
It was reported that the hole density is only 15-30\,\% of the 
total concentration of the Mn atoms.\cite{santos}

\begin{table}[bth]
    \begin{tabular}{{p{1cm}p{1.4cm}p{2cm}p{1.5cm}}}
      \\[-3pt]
    \hline
    \hline
    \\[-8pt]
    \ \ 
    &$T_c$& $M_s$& $N_{\rm Mn}$\\
    \ \ &\hspace{-0.05cm}(K) &
    \hspace{-0.5cm}(emu$\cdot$cm$^{-3}$)&\hspace{-0.5cm}($10^{20}$\,cm$^{-3}$)\\[1pt]
    \hline
    \\[-7pt]
    A$^a$ &130 &38 &8 \\
    \hline
    \\[-7pt]
    B$^a$ &157 &47 &10 \\
    \hline
    \\[-7pt]
    C$^b$ &114 &33 &6.9 \\[1pt]
    \hline
    \\[-7pt]
    D$^c$ &110 & -- &-- \\[1pt]
    \hline
    \\[-7pt]
    E$^d$ &139 &53.5 &11.5 \\[1pt]
    \hline
    \hline
    \\[-3pt]
    \mbox{$^a$ Ref.\ \onlinecite{khazen},\hspace{0.25cm}
      $^b$ Ref.\ \onlinecite{adam},\hspace{0.25cm}
      $^c$ Ref.\ \onlinecite{sinova},\hspace{0.25cm}
      $^d$ Ref.\ \onlinecite{haghgoo}}
  \end{tabular}
  \caption{
    The parameters obtained from the experiments for different samples:
    A: Ga$_{0.93}$Mn$_{0.07}$As/Ga$_{0.902}$In$_{0.098}$As;
    B: Ga$_{0.93}$Mn$_{0.07}$As/GaAs;
    C: Ga$_{0.93}$Mn$_{0.07}$As/Ga$_{1-y}$In$_{y}$As; 
    D: Ga$_{0.92}$Mn$_{0.08}$As;
    E: Ga$_{0.896}$Mn$_{0.104}$As$_{0.93}$P$_{0.07}$.
    $M_s$ stands for the saturate 
    magnetization at zero temperature $M(0)$.
  }
  \label{table1}
\end{table}

In our calculation, the magnetization lies along the principle axis chosen as
[001]-direction.\cite{cywinski}
The conventional parameters are mainly taken from those of GaAs in
Refs.\,\onlinecite{madelung} and \onlinecite{winkler}. Other sample-dependent
parameters such as the Curie temperature and effective Mn concentration are
picked up from the experimental works.\cite{khazen,adam,sinova,haghgoo}
For sample A, B and E (C), only the saturate magnetization at 4 (104)~K was given
in the references. Nevertheless, one can extrapolate the zero temperature
magnetization $M_s$ from 
Eq.\,(\ref{eq3}). The effective Mn concentrations listed in Table\,\ref{table1} are
derived from $N_{\rm Mn}=M_s/(g\mu_B S_d)$. It is clear to see that all of these
effective Mn concentrations are much smaller than the doping density ($\ge1.5\times
10^{21}$~cm$^{-3}$) due to the compensation effect as discussed above. Since the
saturate magnetization of sample D is unavailable, we treat the effective Mn
concentration as a parameter in this case.
Moreover, since the exact values of the itinerant hole densities are
unclear in such strongly disordered samples, we treat them as parameters.
Two typical values are chosen in our numerical calculation, i.e., $N_h=3\times
10^{20}$~cm$^{-3}$ and $5\times 10^{20}$~cm$^{-3}$. The effective impurity
density is taken to be equal to the itinerant hole density.

\begin{figure}[bth]
\includegraphics[width=8.5cm]{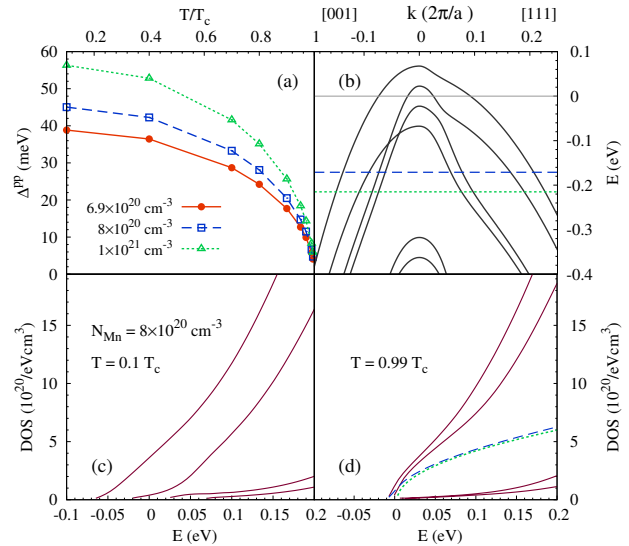}
\caption{(Color online) (a) Zeeman energy as function of temperature. (b) The
  valence band structure with $\Delta^{pp}=45$~meV. The blue dashed curve illustrates
  the Fermi level for the hole density $N_h=3\times 10^{20}$~cm$^{-3}$, while
  the green dotted one gives $N_h=5\times 10^{20}$~cm$^{-3}$.
  The density of states as function of energy at (c) $T/T_c=0.1$ and (d)
  $T/T_c=0.99$ for the uncompensated Mn density $N_{\rm
    Mn}=8\times 10^{21}$~cm$^{-3}$. In (d), the blue dashed curve stands for the
  upper heavy hole band from the spherical approximation and the
  corresponding DOS from the analytical formula
  ($\sqrt{2E}[\sqrt{m^\ast}/(2\pi\hbar)]^3$) is given as the green dotted 
  curve. Here, $J_{\rm ex}^{pp}=-0.5$~eV.
} 
\label{fig1}
\end{figure}

For numerical calculation of the hole spin dynamics, the momentum space is partitioned into
blocks. Compared to the isotropic parabolic dispersion, the band
structure in ferromagnetic GaMnAs is much more complex as we mentioned above
[referred to Figs.\,\ref{fig1}(b) and \ref{fig4}]. Therefore, we need to extend
the partition scheme used in 
isotropic parabolic dispersion\cite{jiang} into anisotropic case.
In our scheme, the radial partition is still carried out with respect to the equal-energy
shells, while the angular partition is done by following
Ref.\,\onlinecite{jiang}. In contrast to the isotropic case, the number of
states in one block is generally different from that in another block even both of them
are on the same equal-energy shell. We calculate the number of states of each
block from its volume in momentum space.

\subsection{Density of states}
By solving Eq.\,(\ref{eq3}), one obtains the magnetization at
finite temperature $M(T)$ and the corresponding Zeeman energy
$\Delta^{pp}$. In Fig.\,\ref{fig1}(a), the Zeeman splitting from $J_{\rm
  ex}^{pp}=-0.5$~eV is plotted as function of the temperature. It is seen
that the Zeeman energy is tens of milli-electron volts at low temperature and
decreases sharply near the Curie temperature due to the decrease of the
magnetization. 
To show the anisotropic nonparabolic feature of the band structure in the presence of the
Zeeman splitting, we illustrate the valence bands along [001]- and [111]-directions in
Fig.\,\ref{fig1}(b), which are obtained from Eq.\,(\ref{eq4}) at $T/T_c=0.1$ for
$N_{\rm Mn}=8\times 10^{20}$~cm$^{-3}$. In this case, the Zeeman splitting
$\Delta^{pp}=45$~meV. The Fermi levels for the hole densities $N_h=3\times
10^{20}$~cm$^{-3}$ and $5\times 10^{20}$~cm$^{-3}$ are shown as blue dashed
and green dotted curves, respectively. As one can see that all of the
four upper bands can be occupied and the effective mass approximation
obviously breaks down. 

By integrating over the volume of each
equal-energy shell, one obtains the density of states (DOS) of each band as
function of energy in Fig.\,\ref{fig1}(c) and
(d). Here the energy is defined in the hole 
picture so that the sign of the energy is opposite to that in
Fig.\,\ref{fig1}(b). It is seen that the DOS of the upper two bands are much
larger than those of the other bands, regardless of the magnitude of the Zeeman
splitting. For $T/T_c=0.99$, the systems approaches the paramagnetic 
phase and the nonparabolic effect is still clearly seen from the DOS in
Fig.\,\ref{fig1}(d), especially in the high energy regime. Moreover, the
pronounced discrepancy of the DOS for the two heavy hole bands suggests the
finite splitting between them. We find
that these features are closely connected
with the anisotropy of the valence bands, corresponding to the Luttinger parameters
$\gamma_2\ne\gamma_3$ in GaAs.\cite{luttinger} In our calculation,
we take $\gamma_1=6.85$, $\gamma_2=2.1$ and $\gamma_3=2.9$ from
Ref.\,\onlinecite{winkler}. As a comparison, we apply a spherical approximation
($\gamma_1=6.85$ and $\gamma_2=\gamma_3=\bar\gamma=2.5$) and find that the
two heavy hole bands become approximately degenerate.\cite{shen3} The DOS of the
upper heavy hole
band is shown as the blue dashed curve in Fig.\,\ref{fig1}(d), where we also
plot the corresponding DOS from the analytical expression, i.e.,
  $\sqrt{2E}[\sqrt{m^\ast}/(2\pi\hbar)]^3$, as the green dotted curve.
Here, we use the heavy-hole effective mass
$m^\ast=m_0/(\gamma_1-2\bar\gamma)$ with $m_0$ denoting the free electron mass.
The perfect agreement between the analytical and our numerical results under the
spherical approximation suggests the good precision of our numerical scheme.

\begin{figure}[bth]
\includegraphics[width=6.5cm]{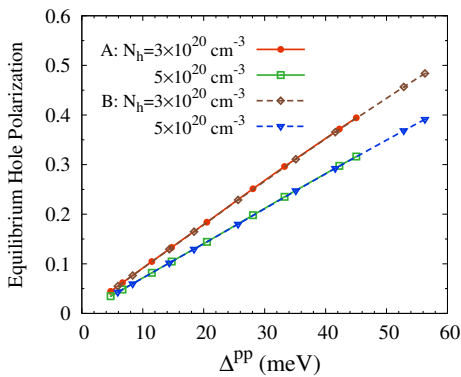}
\caption{(Color online) The equilibrium hole spin polarization as function of
  Zeeman splitting for sample A and B. Here, $J_{\rm ex}^{pp}=-0.5$~eV.
} 
\label{fig2}
\end{figure}

\subsection{Hole spin relaxation}
In this part, we investigate the hole spin dynamics by numerically solving
the microscopic kinetic equation, i.e., Eq.\,(\ref{eq5}). By taking into account the
equilibrium hole spin
polarization, we fit the temporal evolution of the total spin
polarization along [001]-direction by
\begin{equation}
  {\cal J}_z(t)={\cal J}_z^0+{\cal J}_z^\prime e^{-t/\tau_s},
\label{eqfit}
\end{equation}
where ${\cal J}_z^0$ and ${\cal J}_z^\prime$ correspond to the equilibrium and
non-equilibrium spin polarizations, respectively. $\tau_s$ is the
hole spin relaxation time.

In all the cases of the present work, the equilibrium hole spin polarization
for a fixed hole density is found to be approximately linearly
dependent on the Zeeman splitting.
In Fig.\,\ref{fig2}, ${\cal
  J}_z^0$ in samples A and B (similar behavior for others) are plotted as
function of Zeeman splitting, where the exchange coupling constant $J_{\rm
  ex}^{pp}$ is taken to be $-0.5$~eV. One notices that the average spin
polarization becomes smaller with 
the increase of the hole density, reflecting the large interband mixing for the
states in the high energy regime.

\begin{figure}[bth]
\includegraphics[width=6.5cm]{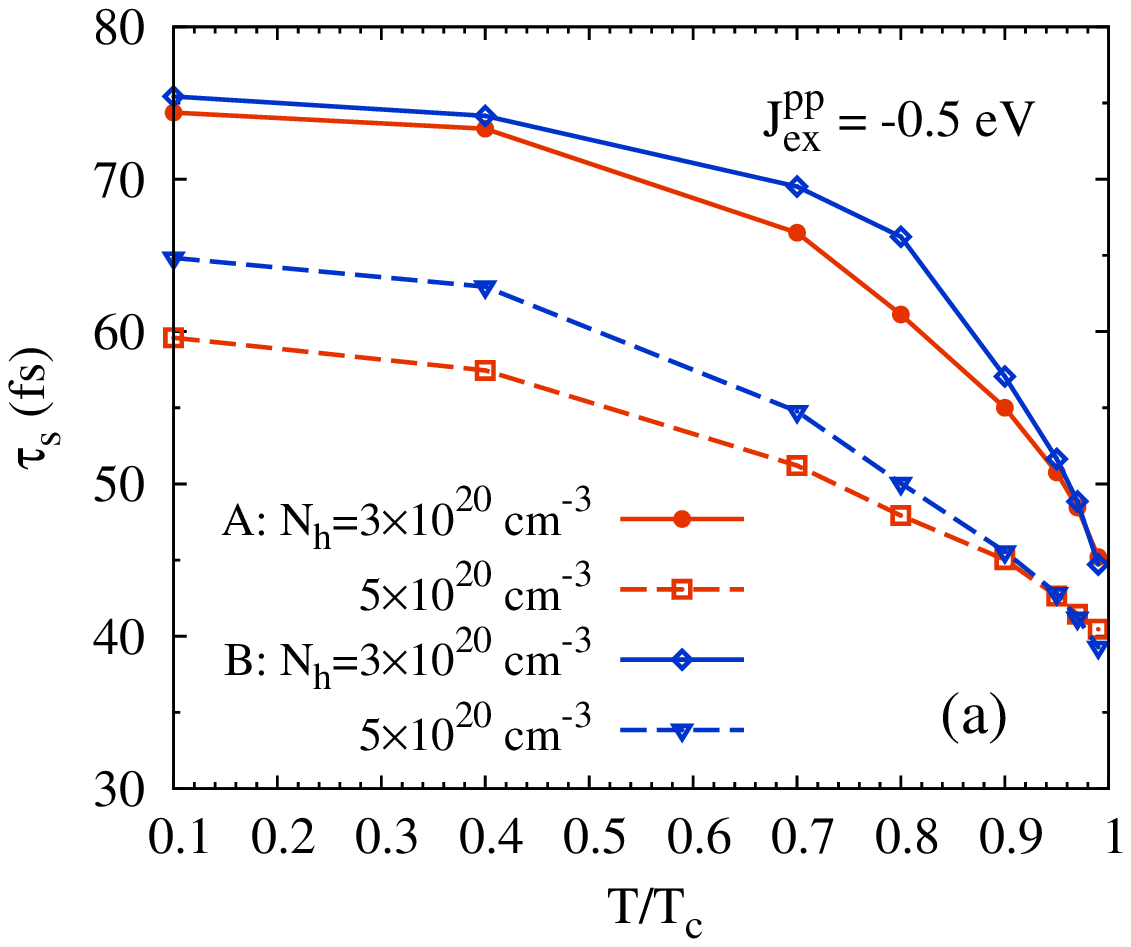}
\includegraphics[width=6.5cm]{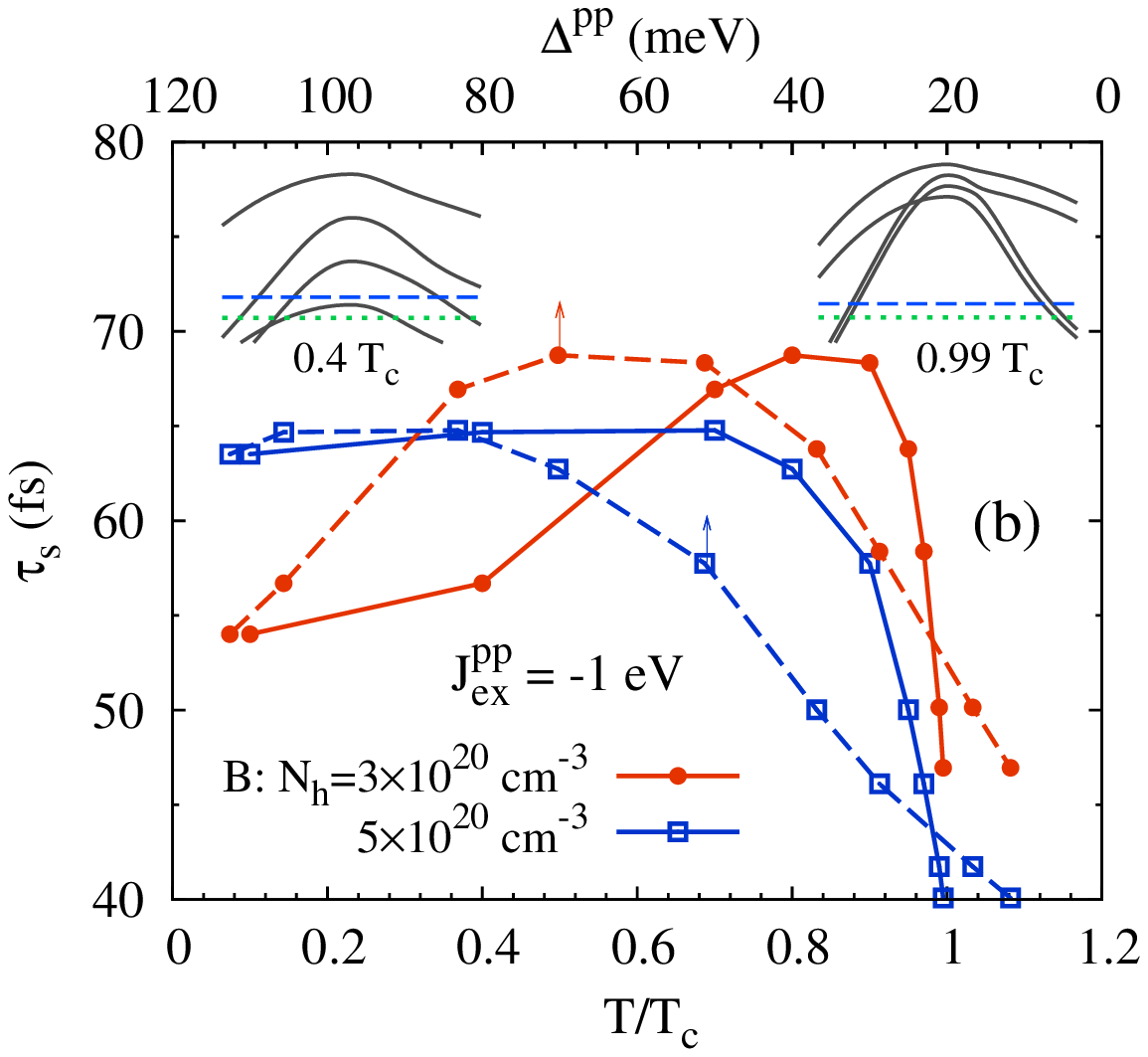}
\caption{(Color online) (a) Spin relaxation time as function of temperature with
  $J_{\rm ex}^{pp}=-0.5$~eV for sample A and B. (b) Spin relaxation time
  as function of temperature and Zeeman splitting obtained from the calculation
  with $J_{\rm ex}^{pp}=-1$~eV for sample B. The inset at the left
  (right) upper corner illustrates the band structure from [001]-direction to
  [111]-direction [refer to Fig.\,\ref{fig1}(b)] for
  $T/T_c=0.4$ (0.99) and $\Delta^{pp}=105$~(16.7)~meV. The
  Fermi levels of $N_h=3\times 10^{20}$~cm$^{-3}$ and $5\times
  10^{20}$~cm$^{-3}$ are shown as the blue dashed and green dotted curves
  in the insets, separately.
}
\label{fig3}
\end{figure}

The temperature dependence of the hole spin relaxation time in samples A and B
with $J_{\rm ex}^{pp}=-0.5$~eV is shown in Fig.\,\ref{fig3}(a),
where the spin relaxation time monotonically decreases with increasing
temperature. This feature can be understood from the
enhancement of the interband mixing as the Zeeman splitting
decreases (shown below).\cite{phonon} 
To gain a complete picture of the
role of the Zeeman splitting on the hole spin relaxation in ferromagnetic
GaMnAs, we also carry out the calculation with the exchange constant $J_{\rm
  ex}^{pp}=-1$~eV.\cite{burch,cywinski} Very interestingly, one finds that the hole spin
relaxation time at low temperature increases with increasing temperature,
resulting in a nonmonotonic temperature dependence 
of the hole spin relaxation time in sample B.
The results in this case are shown as solid curves in
Fig.\,\ref{fig3}(b), where we also plot the Zeeman splitting dependence of the
hole spin relaxation time as
dashed curves. It is seen that the hole spin relaxation time for the hole
density $N_h=3\times 10^{20}$~cm$^{-3}$ first increases with increasing
temperature (alternatively speaking, decreasing Zeeman splitting) and starts to
decrease at around $0.8$~$T_c$ where the Zeeman splitting $\Delta^{pp}=70$~meV.
To understand this feature, we show the typical band
structure in the increase (decrease) regime of the hole relaxation time at
$T/T_c=0.4$~(0.99), corresponding to 
$\Delta^{pp}=105$~(16.7)~meV, in the inset at the left (right) upper corner.
The Fermi levels of the hole density $3\times
10^{20}$~cm$^{-3}$ are labeled by blue dashed
curves. One finds that the carrier occupations in the increase and decrease
regimes are quite different. Specifically, all of the four upper bands are
occupied in the decrease regime while 
only three valence bands are relevant in the
increase regime. 

One may naturally expect that the increase regime originates from
the contribution of the fourth band via the inclusion of the additional
scattering channels or the
modification of the screening. However, we rule out this possibility
through the computation with the fourth band artificially excluded, where the
results are qualitatively the same as those in Fig.\,\ref{fig3}(b).
Moreover, the variations of the screening and the 
equilibrium distribution at finite
temperature are also demonstrated to be irrelevant to the present nonmonotonic
dependence by our calculation (not shown here). Therefore, the interesting
feature has to be attributed to the variations of the
band distortion and spin mixing due to the exchange interaction. 
This is supported by our numerical calculation, where the nonmonotonic
behavior disappears once the effect of the interband mixing is excluded by
removing the wave-function overlaps $|\langle \mu{\bf k}|\nu{\bf k}^\prime\rangle|^2$
in Eqs.\,(\ref{eq6}) and (\ref{eq7}) (not shown here).

\begin{figure}[bth]
\includegraphics[width=8.5cm]{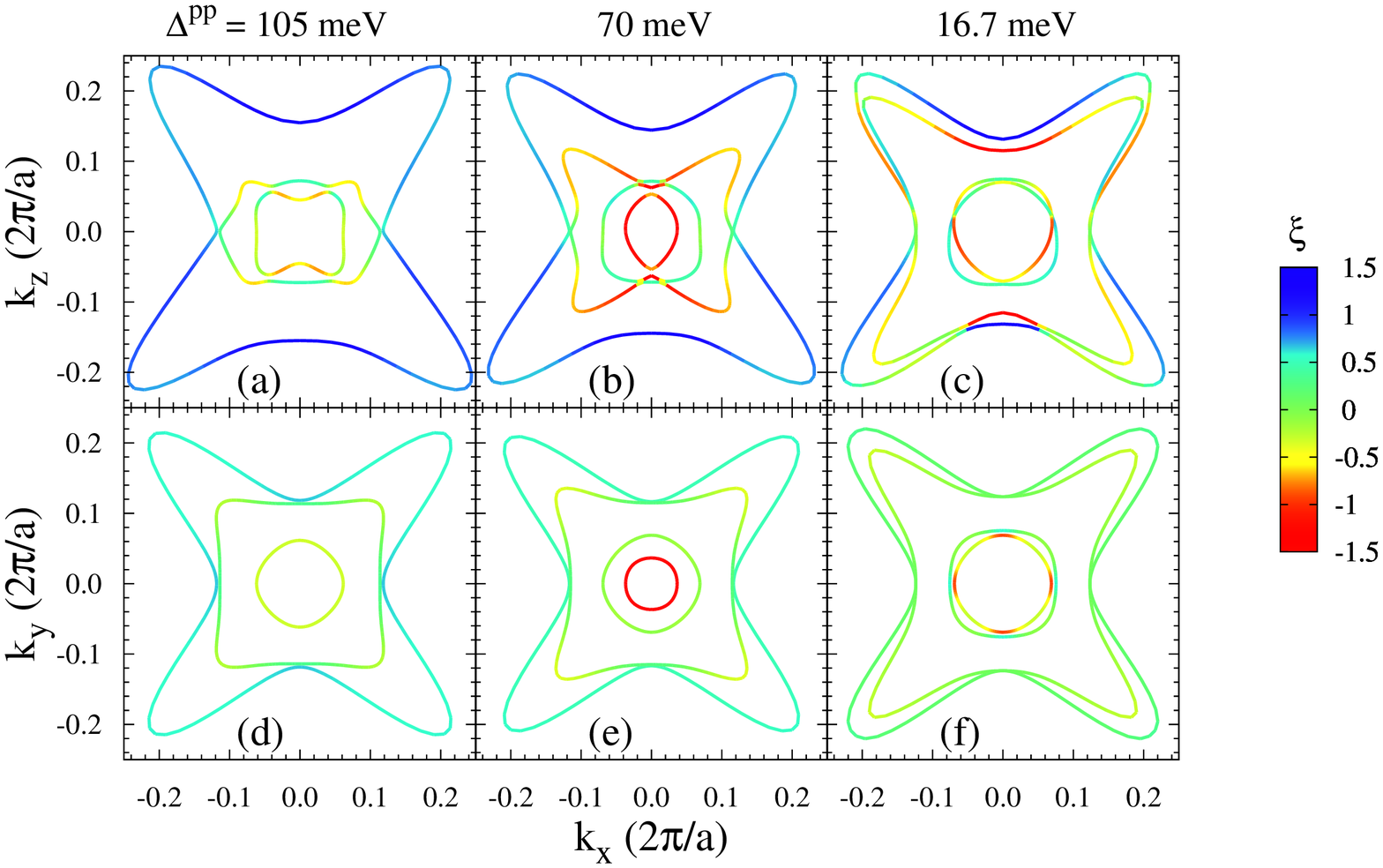}
\caption{(Color online) The Fermi surface in the $k_x$-$k_z$ ($k_y=0$) and
  $k_x$-$k_y$ ($k_z=0$) planes with $\Delta^{pp}$=105~meV
  (a,d), 70~meV (b,e) and
  16.7~meV (c,f). The color coding represents the spin expectation
  of each state, $\xi=\langle \mu|{\cal J}_z|\mu\rangle$. Here, $N_h=3\times
  10^{20}$~cm$^{-3}$.
}
\label{fig4}
\end{figure} 

For a qualitative understanding of the nonmonotonic temperature dependence
  of the hole spin relaxation time,
we plot the Fermi surface in the $k_x$-$k_z$ ($k_y=0$) and
$k_x$-$k_y$ ($k_z=0$) planes at $N_h=3\times 10^{20}$~cm$^{-3}$ in
Fig.\,\ref{fig4}. We choose typical Zeeman splittings in the increase regime
($\Delta^{pp}=105$~meV), the decrease regime ($\Delta^{pp}=16.7$~meV)
 and also the crossover regime ($\Delta^{pp}=70$~meV). 
One notices that the Fermi surfaces in Fig.\,\ref{fig4}(a) and
(d) are composed of three closed curves, meaning that only three bands
are occupied for $\Delta^{pp}=105$~meV [also see the inset of Fig.\,\ref{fig3}(b)].
For the others with smaller Zeeman splittings, all of the four upper bands are occupied.
The spin expectation of each state at
Fermi surface is represented by the color coding. 
Note that the spin expectation
of the innermost band for $\Delta^{pp}=70$~meV
is close to $-1.5$ [see Fig.\,\ref{fig4}(b) and (e)], suggesting that this band is the
spin-down heavy hole band and the mixing of other spin components in this band is marginal.
Therefore, the spin-flip scattering related to this band is weak and can
not result in the increase of the hole spin relaxation time mentioned above.
By comparing the results with $\Delta^{pp}=105$~meV and $70$~meV, one notices
that the spin expectation of the Fermi surface of the outermost band is
insensitive to the Zeeman splitting. Therefore, this band can not be the reason of
the increase regime also. Moreover, for the second and third
bands in Fig.\,\ref{fig4}(a) and (b), from the comparable color coding between
the two figures in this regime [also see Fig.\,\ref{fig4}(d) and (e) with
$k_z=0$], one finds that the spin 
expectation for the states with small $k_z$ is also insensitive to the Zeeman splitting.
However, for the states with large $k_z$, the spin expectation of the spin-down
states ($\xi<0$) approaches a large magnitude ($-1.5$) with decreasing Zeeman 
splitting, suggesting the decrease of the mixing from the spin-up states.
As a result, the interband spin-flip scattering from/to these states becomes weak and
the hole spin relaxation time increases. 
In the decrease regime of the hole spin relaxation time, Fig.\,\ref{fig4}(c) and
(f) show that the two outer/inner bands approach each other,
leading to a strong and anisotropic spin mixing.
Therefore, the spin-flip scattering becomes more efficient in this regime
and the spin relaxation time decreases. One may suppose that the nonmonotonic
temperature dependence of the hole spin relaxation time can also arise from the
variation of the shape of the Fermi surface, according to
Fig.\,\ref{fig4}. However, this
variation itself is not the key of the nonmonotonic behavior, because the
calculation with this effect but without band mixing can not recover the
nonmonotonic feature as mentioned in the previous paragraph.
For the hole density $N_h=5\times 10^{20}$~cm$^{-3}$, the structures
of the Fermi
surface at $\Delta^{pp}=105$~meV are similar to those in Fig.\,\ref{fig4}(b) and
(e). This explains the absence of the increase regime for this density in
Fig.\,\ref{fig3}(b).

 Moreover, we should point out that the increase
regime of the hole spin relaxation time in sample A for
$J_{\rm ex}^{pp}=-1$~eV is much narrower than that in sample B. The reason lies
in the fact of lower effective Mn density in sample A, leading to the smaller
maximal Zeeman splitting $\sim 90$~meV, only slightly larger than the crossover
value 70~meV at $N_h=3\times 10^{20}$~cm$^{-3}$.

As a summary of this part, we find different temperature dependences of
the hole spin relaxation time due to the different values of effective Mn
concentration, hole density and  exchange coupling constant
$J_{\rm ex}^{pp}$. In the case with 
large coupling constant and high effective Mn concentration, the interband spin mixing can
result in a nonmonotonic temperature dependence of the hole spin relaxation time. Our
results suggest a possible way to estimate the exchange coupling constant with the
knowledge of itinerant hole density, i.e.,
by measuring the temperature dependence of the hole spin
relaxation time. Alternatively, the discrepancy between the hole relaxation time
from different hole densities in Fig.\,\ref{fig3}(b) suggests that one can also
estimate the itinerant hole
density if the exchange coupling constant has been measured from other methods.

\subsection{Gilbert damping and non-adiabatic torque coefficients}
Facilitated with the knowledge of the hole spin relaxation time, we can calculate the
coefficients in the LLG equation. According to our previous works,\cite{shen1,shen2} 
the Gilbert damping and nonadiabatic spin torque coefficients can be expressed as
\begin{equation} 
\alpha={J_h}/[N_{\rm Mn}{|\langle {\bf S}\rangle|(\beta+1/\beta)}],
\label{eqalpha}
\end{equation}
and
\begin{equation}
  \beta={1}/{(2\tau_s
  \Delta^{pp})},
\label{eqbeta}
\end{equation} 
respectively. In Eq.\,(\ref{eqalpha}), $J_h$ represents the total equilibrium
spin polarization of the itinerant hole gas, i.e.,
$J_h=N_h{\cal J}_z^0$ with ${\cal J}_z^0$ being the one defined in Eq.\,(\ref{eqfit})
in our study. The average spin polarization of a single Mn ion is
given by $|\langle {\bf S}\rangle|=S_dM(T)/M(0)$.

\begin{figure}[bth]
\includegraphics[width=8.5cm]{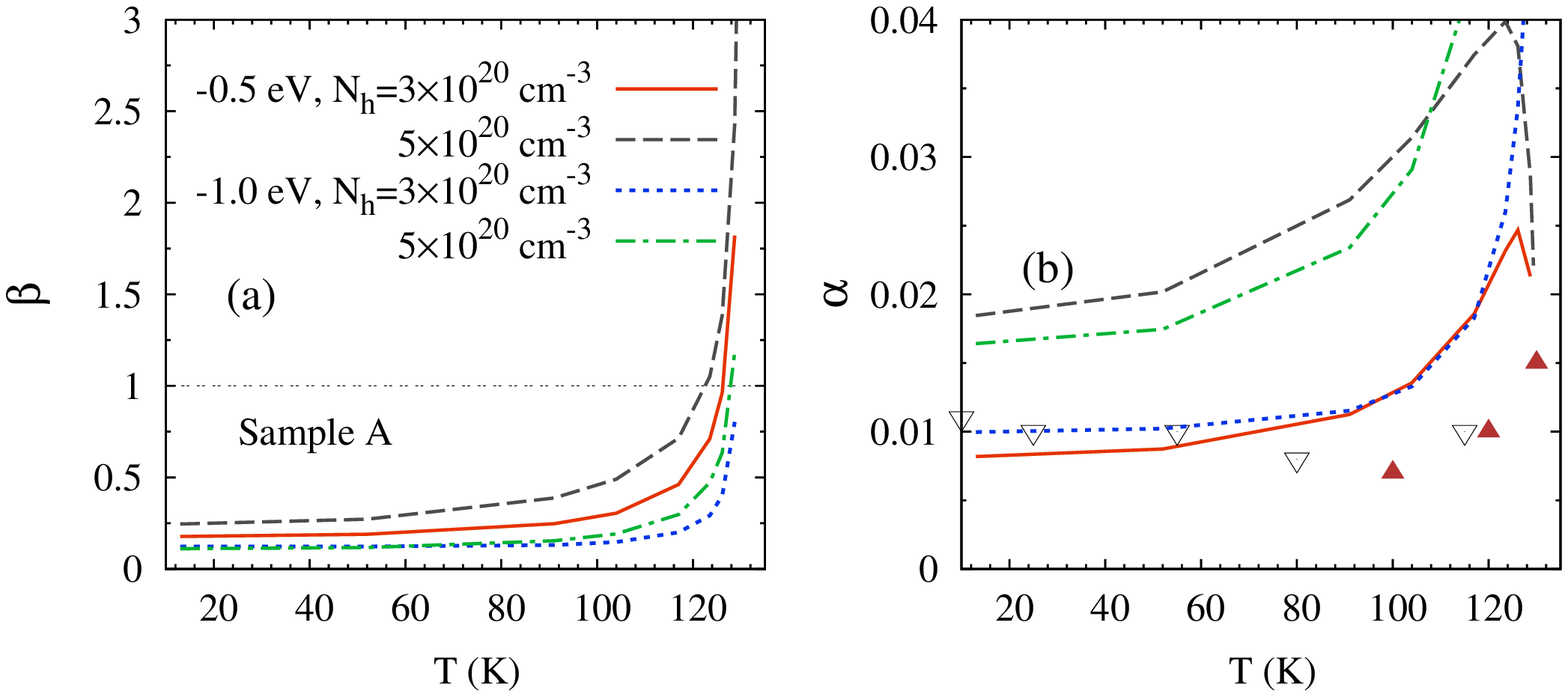}
\vskip -0.4cm
\includegraphics[width=8.5cm]{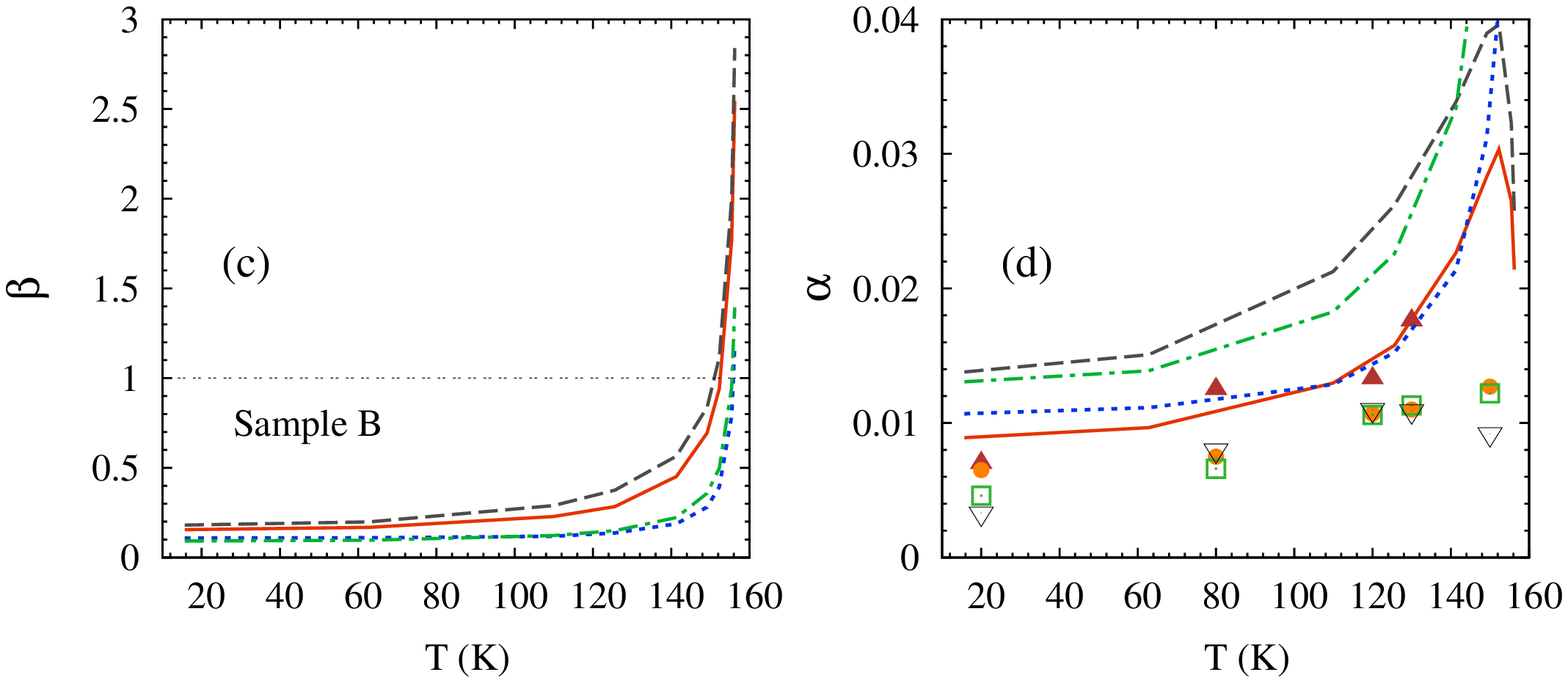}
\vskip -0.4cm
\includegraphics[width=8.5cm]{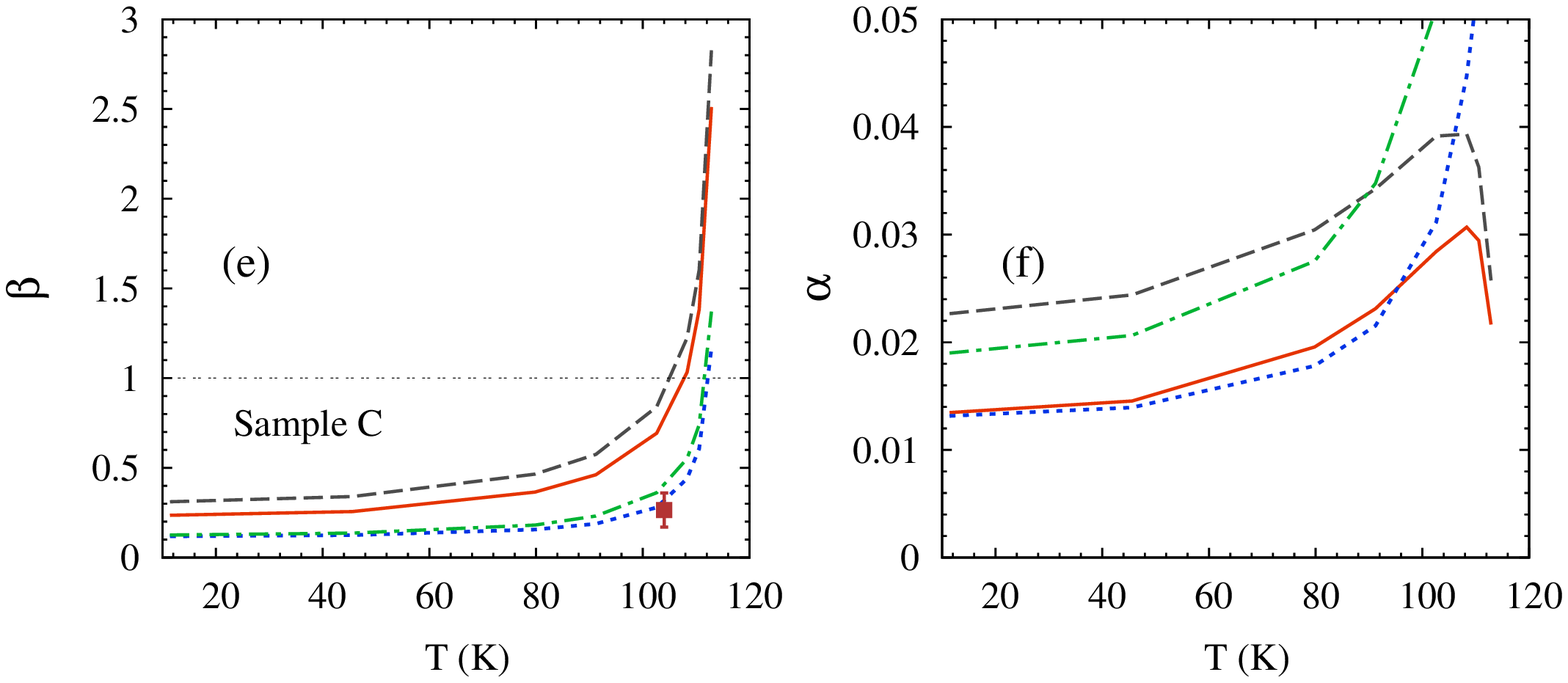}
\caption{(Color online) $\beta$ and $\alpha$ as function of temperature
  with $J_{\rm ex}^{pp}=-0.5$~eV and $-1.0$~eV in sample A-C. In (b) and (d),
  the dots represent the experimental data from ferromagnetic resonance measurement for
  [001] (brown solid upper triangles), [110] (orange solid circles), [100]
  (green open squares) and [1-10] (black open lower triangles) dc magnetic-field orientations
  (Ref.\,\onlinecite{khazen}). The brown solid square in (e) stands for the
  experimental result from domain-wall motion measurement (Ref.\,\onlinecite{adam}).
}
\label{fig5}
\end{figure}

In Fig.\,\ref{fig5}(a), (c) and (e), the nonadiabatic spin torque
coefficients $\beta$ in sample A-C are plotted as function of
temperature with $J_{\rm ex}^{pp}=-0.5$~eV and $-1.0$~eV.
Our results in sample C show good agreement with the experimental data (plotted
as the brown square) in Fig.\,\ref{fig5}(e).\cite{adam}
At low temperature, the value of $\beta$ is around 0.1$\sim$0.3, which is
also comparable with the previous theoretical calculation.\cite{garate}  Very
interestingly, one finds that $\beta$ sharply increases when the temperature
approaches the Curie 
temperature. This can be easily understood from the pronounced decreases of
the spin relaxation time and the Zeeman splitting in this regime [see Figs.\,\ref{fig1}(a) and
\ref{fig3}]. By comparing the results with different values of the exchange
coupling constant,
one finds that $\beta$ from $J_{\rm ex}^{pp}=-1$~eV is generally about one half of that
obtained from $J_{\rm ex}^{pp}=-0.5$~eV because of the larger Zeeman
splitting. Moreover, one notices that the nonmonotonic temperature dependence of
the hole spin relaxation time in Fig.\,\ref{fig3}(b) is not reflected in $\beta$
due to the influence of the Zeeman splitting. In all cases, the values of $\beta$ can
exceed one very near the Curie temperature.

The results of the Gilbert damping coefficient from Eq.\,(\ref{eqalpha}) are
shown as curves in Fig.\,\ref{fig5}(b), (d) and (f). The dots in these figures
are the reported experimental data from the ferromagnetic resonance along different
magnetic-field orientations.\cite{khazen} Both the magnitude and the temperature
dependence of our results agree well with the experimental data. From Fig.\,\ref{fig2}, one
can conclude that the prefactor in Eq.\,(\ref{eqalpha}),
$J_h/(N_{\rm Mn}|\langle{\bf S}\rangle|)$, is almost independent of
temperature. Therefore, the temperature dependence of $\alpha$ mainly results from
the nonadiabatic spin torque coefficient $\beta$. Specifically,
$\alpha$ is insensitive to the temperature in the low temperature
regime and it
gradually increases with increasing temperature due to the increase of $\beta$.
Moreover, we predict that $\alpha$ begins to decrease with increasing temperature
once $\beta$ exceeds one. This crossover lying at
$\beta\approx 1$ can be expected from Eq.\,(\ref{eqalpha}). By comparing the
results with different values of $J_{\rm ex}^{pp}$, one finds that the
value of $\alpha$ is robust against the exchange coupling constant in the low
temperature regime. In this regime, $\beta\ll1$ and one can simplify the
expression of the Gilbert damping coefficient as $\alpha\approx
\frac{N_h}{N_{\rm Mn}S_d}\frac{{\cal
  J}_z^0}{(\tau_s\Delta^{pp})}$. Since 
the total hole spin polarization is proportional to the Zeeman splitting
(see Fig.\,\ref{fig2}) and $\tau_s$ is only weakly dependent on the Zeeman
splitting (see Fig.\,\ref{fig3}) in this regime, the increase of $J_{\rm ex}^{pp}$
does not show significant effect on $\alpha$. However, at high temperature, the scenario
is quite different. For example, one has the maximum of the Gilbert damping coefficient
$\alpha_m\approx \frac{N_h}{2N_{\rm Mn}|\langle{\bf S}\rangle|}{{\cal J}_z^0}\propto J_{\rm
  ex}^{pp}$ at $\beta=1$.

\begin{figure}[bth]
\includegraphics[width=8.5cm]{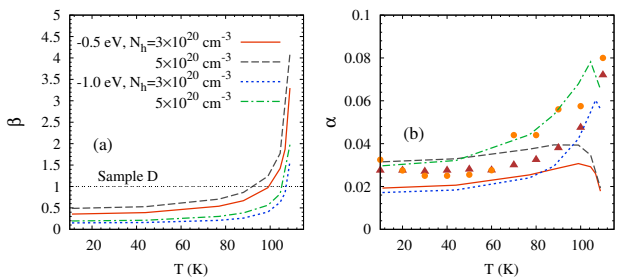}
\caption{(Color online) $\beta$ and $\alpha$ as function of temperature by
  taking $N_{\rm Mn}=5\times 10^{20}$~cm$^{-3}$ with $J_{\rm ex}^{pp}=-0.5$~eV and
  $-1.0$~eV in sample D. The dots are from ferromagnetic resonance measurement
  (Ref.\,\onlinecite{sinova}) for [001] (brown solid 
  upper triangles) and [110] (orange solid circles) dc magnetic-field orientations.
}
\label{fig6}
\end{figure}

Since the effective Mn concentration of sample D is unavailable as mentioned
above, we here take $N_{\rm Mn}=5\times 10^{20}$~cm$^{-3}$. The results
are plotted in Fig.\,\ref{fig6}. It is seen that the Gilbert damping coefficients
from our calculation with $J_{\rm ex}^{pp}=-1$~eV agree with the experiment
very well. As reported, the 
damping coefficient in this sample is much larger ($\sim 0.1$) before annealing.\cite{sinova}
The large Gilbert damping coefficient in the as-grown sample may result from
the direct spin-flip scattering between the 
holes and the random Mn spins, existing in low quality samples. In the
presence of this additional spin-flip channel, the hole spin relaxation time
becomes shorter and results in an enhancement of $\alpha$ and $\beta$ (for $\beta<1$).
Moreover, in the low temperature regime, a decrease of the Gilbert
damping coefficient was observed by increasing
temperature,\cite{sinova} which is absent in our
results. This may originate from the complicated localization or correlation
effects in such a disordered situation. The quantitatively microscopic study in
this case is beyond the scope of the present work.

In addition, one notices that $\beta$ in Ref.\,\onlinecite{yaman} was
determined to be around 0.01,
which is one order of magnitude smaller than our result. The reason is because of the
incorrect parameter used in that work, as pointed out by Adam {\em et
    al.}.\cite{adam} 

\begin{figure}[bth]
\includegraphics[width=7.cm]{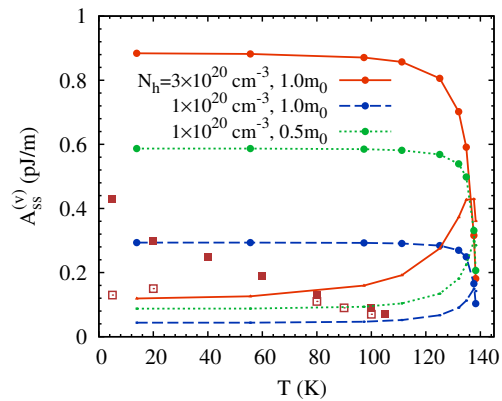}
\caption{(Color online) Spin stiffness (vertical spin stiffness) coefficient as function of
  temperature is plotted as curves with (without) symbols. The calculation
  is carried out with $J_{\rm ex}^{pp}=-0.5$~eV in sample E.
  The effective mass is taken to be 1.0 (0.5)$m_0$ as labeled in the figure.
  The brown solid (from the period of the domains) and open
  (from the hysteresis cycle) squares are the experimental data of spin
  stiffness from Ref.\,\onlinecite{haghgoo}.  
}
\label{fig7}
\end{figure}

\subsection{Spin stiffness and vertical spin stiffness}
In this subsection, we calculate the spin stiffness and vertical
spin stiffness coefficients according to our previous derivation\cite{shen2}
\begin{equation}
  A_{\rm ss}={N_h}/[{4m^\ast(1+\beta^2)}]
\label{eq15}
\end{equation}
and
\begin{equation}
  A_{\rm ss}^{\rm v}={N_h}\beta/[{4m^\ast(1+\beta^2)}].
\label{eq16}
\end{equation}
Since the effective mass $m^\ast$ is a rough description for the
anisotropic valence bands in the presence of a large Zeeman splitting, it
is difficult to obtain the accurate
value of the stiffness coefficients from these formulas. Nevertheless,
one can still estimate these coefficients with the effective mass taken as a parameter.
The results are plotted in Fig.\,\ref{fig7}. By fitting the DOS of the
  occupied hole states, we find $m^\ast\approx m_0$, which is consistent with
the previous work.\cite{cywinski}
 The spin stiffness and vertical spin
stiffness coefficients with $N_h=3\times 10^{20}$~cm$^{-3}$ ($1\times
  10^{20}$~cm$^{-3}$)
are plotted as the red solid (blue dashed)
curves with and without symbols, respectively.
The sudden decrease of $A_{\rm
  ss}$ originates from the increase of $\beta$ in the vicinity of the Curie
temperature (see Fig.\,\ref{fig5}). Our results are comparable with
the previous theoretical work from 6-band model.\cite{konig2} As a comparison,
we take $m^\ast=0.5m_0$, which is widely used to describe the heavy hole in
  the low energy regime in the absence of the Zeeman splitting.\cite{leo} The spin
stiffness becomes two times larger. Moreover, $A_{\rm ss}^{\rm v}$
is found to present a nonmonotonic behavior in the temperature dependence as
predicted by Eq.\,(\ref{eq16}).

In Fig.\,\ref{fig7}, we also plot the experimental data of the spin stiffness
coefficient from Ref.\,\onlinecite{haghgoo}. It is seen that these values
of $A_{\rm ss}$ are comparable with our results and
show a decrease as the temperature increases. However, one notices that the
experimental data is more sensitive to the temperature especially for those
determined from the domain period in the low temperature regime. 
This may originate from the strong anisotropic interband mixing and
  inhomogeneity in the real material.

In Ref.\,\onlinecite{shen2}, we have shown that the vertical spin
stiffness can lead to the magnetization rotated around
the easy axis within the domain wall structure by $\Delta
\varphi=(\sqrt{1+\beta^2}-1)/\beta$ in the absence of the
demagnetization field. For $\beta= 1$, $\Delta\varphi\approx
0.13\pi$, while $\Delta\varphi=\beta/2\to 0$ for $\beta\ll 1$. As
illustrated above, $\beta$ is always larger than 0.1. Therefore, the vertical
spin stiffness can present observable modification
of the domain wall structure in GaMnAs system.\cite{shen2}

\section{Summary}\label{conclusion}
In summary, we theoretically investigate the temperature dependence of the LLG
coefficients in ferromagnetic GaMnAs, based on the
microscopic calculation of the hole spin relaxation time.
In our calculation, we employ the Zener model with the band
structure carried out by diagonalizing the $8\times 8$ Kane
Hamiltonian together with the Zeeman energy due to the $sp$-$d$ exchange
interaction. We find that the hole spin relaxation time can present different
temperature dependences, depending on the effective Mn concentration, hole
density and exchange coupling constant. In the case with high Mn
concentration and large exchange coupling constant, the hole spin relaxation
time can be nonmonotonically dependent on temperature,
resulting from the different
interband spin mixings in the large and small Zeeman splitting regimes. These features
are proposed to be for the estimation of the exchange coupling constant or
itinerant hole density. By substituting the hole relaxation time, we calculate
the temperature dependence of the Gilbert damping, nonadiabatic spin torque, spin
stiffness, and vertical spin stiffness coefficients. We obtain the
nonadiabatic spin torque coefficient around $0.1\sim 0.3$ at low
temperature, which is consistent with the experiment. As the temperature
increases, this coefficient shows a monotonic increase.
In the low temperature regime, the Gilbert damping
coefficient increases with temperature, which shows good agreement with the experiments.
We predict that the Gilbert damping coefficient can decrease with increasing temperature
once the nonadiabatic spin torque coefficient exceed one in
the vicinity of the Curie temperature. We also find that the spin stiffness
decreases with increasing temperature and the vertical spin stiffness can
present a nonmonotonic temperature dependence, similar to the Gilbert damping.

\begin{acknowledgments}
This work was supported by the National Natural Science Foundation of
China under Grant No.\,10725417 and the
National Basic Research Program of China under Grant No.\,2012CB922002.
\end{acknowledgments}



\begin{thebibliography}{0}
\bibitem{phonon} The hole-phonon scattering is rather weak compared 
with the hole-impurity
 scattering in this low temperature regime. Therefore the pronounced
 temperature dependence of the hole spin relaxation 
time can not be explained by this scattering mechanism.
\end{thebibliography}
\end{document}